%% file: ms.tex
\documentclass[useAMS,usenatbib]{mn2e}

\usepackage{epsfig}
\usepackage{amsmath}
\usepackage{amssymb}
\usepackage{ifthen}
\usepackage{txfonts}
\usepackage{rotating}
\usepackage{url}
\usepackage{varioref}
\usepackage{verbatim}
\usepackage{latexsym}
\usepackage{graphicx}
\usepackage{placeins}
\usepackage{float}
\makeatletter
\def\fps@figure{!htbp}
\makeatother

\DeclareSymbolFont{cmletters}{OML}{cmm}{m}{it}

\DeclareMathSymbol{v}{\mathord}{cmletters}{"76}

\def\be{\begin{equation}}
\def\ee{\end{equation}}

\newcommand{\hrat}{{\bar{h}}}

\newcommand{\abs}[1]{\ensuremath{\left|#1\right|}}

\DeclareSymbolFont{cmletters}{OML}{cmm}{m}{it}
\DeclareMathSymbol{v}{\mathalpha}{cmletters}{"76}

\usepackage{subfigure}
\usepackage{ifthen}
\usepackage[usenames,dvipsnames]{color}
\usepackage{graphicx}

\usepackage{hyperref}


\defcitealias{bz77}{BZ77}

\newcommand{\eff}{{\eta}}

\title[Efficiency of Thin MADs]{Efficiency of Thin Magnetically-Arrested Disks Around Black Holes}
\author[M.~J.~Avara
J.~C.~McKinney and
C.~S.~Reynolds]
{Mark J. Avara$^1$\thanks{\hbox{E-mail: mavara@astro.umd.edu~(MJA)}} ,
Jonathan C. McKinney$^{2,3}$,
Christopher S. Reynolds$^{1,2}$
\\
 $^1$ University of Maryland at College Park, Dept. of Astronomy, 1113 Physical Sciences Complex, College Park, MD 20742, USA
\\
  $^2$Joint Space-Science Institute,1113 Physical Sciences Complex, College Park, MD 27042, USA
\\
  $^3$University of Maryland at College Park, Dept. of Physics, 3114 Physical Sciences Complex, College Park, MD 20742, USA
 }{
}

\begin{document}
\date{Submitted 2015 August xx. xxxxxxxxxxx}
\pagerange{\pageref{firstpage}--\pageref{lastpage}} \pubyear{2015}
\maketitle

\label{firstpage}

\input{msabstract.tex}


\input{msintro.tex}

\input{mssetup.tex}

\input{msdiscuss.tex}

\input{msconclude.tex}

\section*{Acknowledgments}

We thank Ramesh Narayan, Jack Steiner, Roman Gold, Peter Polko,
Alexander Tchekhovskoy, and Eliot Quataert for discussions.  We
acknowledge NASA/NSF/TCAN (NNX14AB46G), NSF/XSEDE/TACC/Stampede
(TG-PHY120005), NASA/Pleiades (SMD-14-5451), and UMD Deepthought2.

%

\bibliographystyle{mnras}
{\small
\bibliography{mybibnew}
}

\label{lastpage}
\end{document}

%% file: msabstract.tex
\begin{abstract}

The radiative and jet efficiencies of thin magnetized accretion disks
around black holes (BHs) are affected by BH spin and the presence of a
magnetic field that, when strong, could lead to large deviations from
Novikov-Thorne (NT) thin disk theory.  To seek the maximum deviations,
we perform general relativistic magnetohydrodynamic (GRMHD)
simulations of radiatively efficient thin (half-height $H$ to radius
$R$ of $H/R\approx 0.10$) disks around moderately rotating BHs with
$a/M=0.5$.  First, our simulations, each evolved for more than $70,000r_g/c$
(gravitational radius $r_g$ and speed of light $c$), show that
large-scale magnetic field readily accretes inward even through our
thin disk and builds-up to the magnetically-arrested disk (MAD) state.
Second, our simulations of thin MADs show the disk achieves a
radiative efficiency of $\eta_{\rm r}\approx 15\%$ (after estimating
photon capture), which is about twice the NT value of $\eta_{\rm
  r}\sim 8\%$ for $a/M=0.5$ and gives the same luminosity as a NT disk
with $a/M\approx 0.9$.  Compared to prior simulations with $\lesssim
10\%$ deviations, our result of an $\approx 80\%$ deviation sets a new
benchmark. Building on prior work, we are now able to complete an important 
 scaling law which suggest that observed
jet quenching in the high-soft state in BH X-ray binaries is
consistent with an ever-present MAD state with a weak yet sustained jet.

\end{abstract}

%% file: msintro.tex
\section{Introduction}
\label{sec:intro}

Black hole (BH) accretion systems can operate as efficient engines
converting gravitational potential energy and BH spin energy into
radiation.  Quasars, active galactic nuclei (AGN), X-ray binaries,
gamma-ray bursts, and other BH accretion systems enter a thin
radiatively efficient mode when the luminosity is between $\sim
10^{-2}$ and $\sim 0.5$ times the Eddington luminosity
\citep{abr95,nar95b}.  The \citet{nov73}, NT, calculation for the thin
disk radiative efficiency assumes emission terminates at the
inner-most stable circular orbit (ISCO).  However, accretion is driven
by magnetic stresses via the magneto-rotational instability (MRI)
\citep{bal91,bh98} or magnetic braking by large scale field threading
the disk \citep{bp82} or BH \citep{bz77}, hereafter BZ.  A fundamental
question in accretion theory has been whether the NT model is 
applicable to actual magnetized disks where magnetic flux couples to a 
rotating BH and increases the radiative efficiency near the ISCO 
\citep{1999ApJ...522L..57G,krolik99}.  Any extra radiative efficiency 
could, in principle, affect BH spin estimates obtained from Soltan's 
argument (e.g., see \citealt{2005ApJ...620...59S}) or from continuum 
emission spectrum observations \citep{mcclintock_spins_2011}. 

An upper limit to the amount of magnetic flux a disk and BH
in steady-state can support is achieved when an
accretion flow reaches the so-called magnetically-arrested disk (MAD)
state, when magnetic flux accretes until magnetic forces pushing out
balance gas forces pushing in
\citep{ina03,2003PASJ...55L..69N,Igumenshchev08,tnm11,2012MNRAS.423.3083M}.
If jet power scales as $P_j\propto B^2 a^2$ for magnetic field $B$ and
BH spin $a/M$ \citep{bz77}, then the MAD state leads to the maximum
jet power.  The existence of such a maximum well-defined magnetic flux
for a given spin is assumed when using transient jets in BH X-ray
binaries to test the BZ prediction of power vs. spin
\citep{2014SSRv..183..295M}.  Destruction of such a strong magnetic
flux could lead to such transient jets without the need for BH spin
\citep{igu09,2014MNRAS.440.2185D}, while BH spin measurements may be
more reliable for thin MADs that could have angular momenta aligned
with the BH spin axis near the BH \citep{2015arXiv151207969P}.

General relativistic magnetohydrodynamic (GRMHD) simulations of thin
disks have found different degrees of deviations from NT theory, but
in these cases there are no more than $10\%$ fractional deviations for
the radiative efficiency
\citep{2008ApJ...687L..25S,2009ApJ...692..411N,2010MNRAS.408..752P,nkh10,2011ApJ...743..115N}. Differences
in results are likely due to the amount of magnetic flux that threads
the BH and disk \citep{2010MNRAS.408..752P}.  Unlike other choices for
initial conditions, the amount of magnetic flux threading the BH and
disk in a MAD state is independent of the initial magnetic field
strength or geometry as long as there is a plentiful supply of
magnetic flux \citep{2012MNRAS.423.3083M,2012MNRAS.423L..55T}.
Numerous non-MAD GRMHD simulations of relatively thick disks have been
performed
\citep{2003ApJ...589..458D,2004ApJ...611..977M,dhkh05,2005ApJ...630L...5M,2006MNRAS.368.1561M,2007MNRAS.377L..49K,2007MNRAS.375..513M,2007ApJ...668..417F,2009MNRAS.394L.126M,2012MNRAS.426.3241N}.
Simulations of MADs have so-far involved only relatively thick disks
with no cooling \citep{tnm11,2012MNRAS.423.3083M} and thick disks with
radiative transfer \citep{2015arXiv150802433M}.

We perform fully 3D GRMHD simulations of radiatively efficient thin
(half-height $H$ to radius $R$ of $H/R\approx 0.1$) disks around
rotating BHs (dimensionless spin, $a/M=0.5$) that reach the MAD state.
We expect the MAD to maximize the radiative efficiency of thin disks
and lead to the maximum deviations from the NT efficiency.  First, we
test how easily magnetic flux advects inward by threading the disk
with weak magnetic field.  Second, we setup a disk that is initially
nearly MAD, let it become MAD to large radius, and then measure the
radiative efficiency.

We describe the physical and numerical setup in
\textsection\ref{sec:setup}, present results and provide discussions
in \textsection\ref{sec:results}, and summarize in
\textsection\ref{sec:summary}.

%% file: mssetup.tex
\section{Fully 3D GRMHD Thin MAD Model}
\label{sec:setup}

For this study we use the HARM GRMHD code \citep{2003ApJ...589..444G}
with various improvements
\citep{2009MNRAS.394L.126M,2012MNRAS.423.3083M} with two physical
setups, both with spin $a/M=0.5$.  Each was run with a low resolution,
medium resolution, and high resolution for convergence testing and the
high resolution was used for detailed analysis.  We name the
simulations in this work MADxy where x=i,f signifies whether the disk
starts with enough flux to be nearly MAD in the initial, i, or only
MAD in the final, f, state.  y=HR,MR distinguishes the grid as high or
medium resolution.

\subsection{Physical Model}

\begin{figure*}
\centering
\includegraphics[width=6.in,clip]{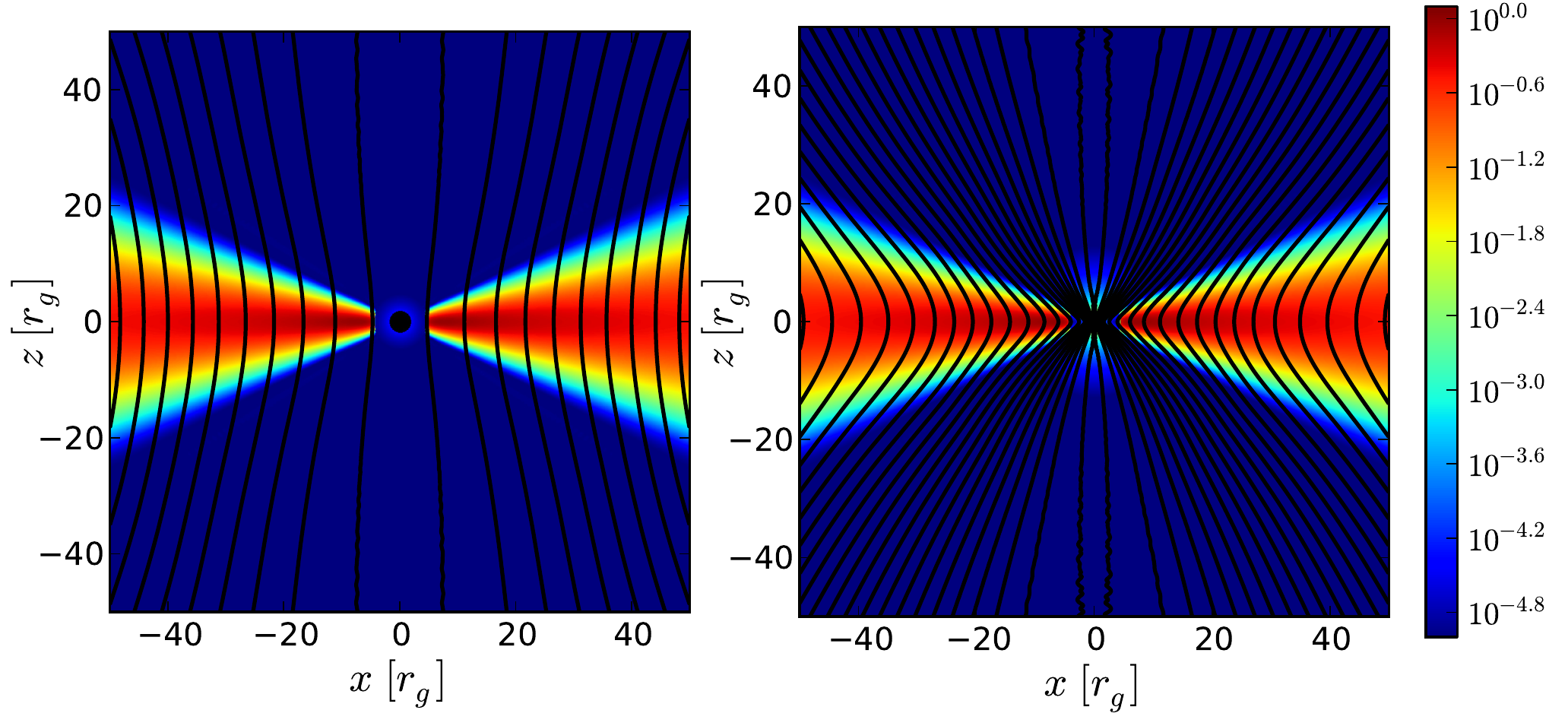}
\caption{Snapshots of the initially sub-MAD (MADfHR - left panel) and
  nearly-MAD (MADiHR - right panel) models at t=0, showing $\rho$ in
  color (with legend) with magnetic field lines (black lines). In
  MADfHR there is no magnetic field initially present within the ISCO
  or on the horizon and the field is weak througout the disk. On the
  other hand, in MADiHR, a sufficient magnetic flux is present
  throughout the disk and on the horizon so that the disk will quickly
  become MAD out to large radius. The higher density surrounding the
  horizon in MADiHR at t=0 is the result of the density floor
  activated due to the presence of an initially strong magnetic field.}
\label{fig:init}
\end{figure*}

\begin{figure*}
\centering
\includegraphics[width=6.3in,clip]{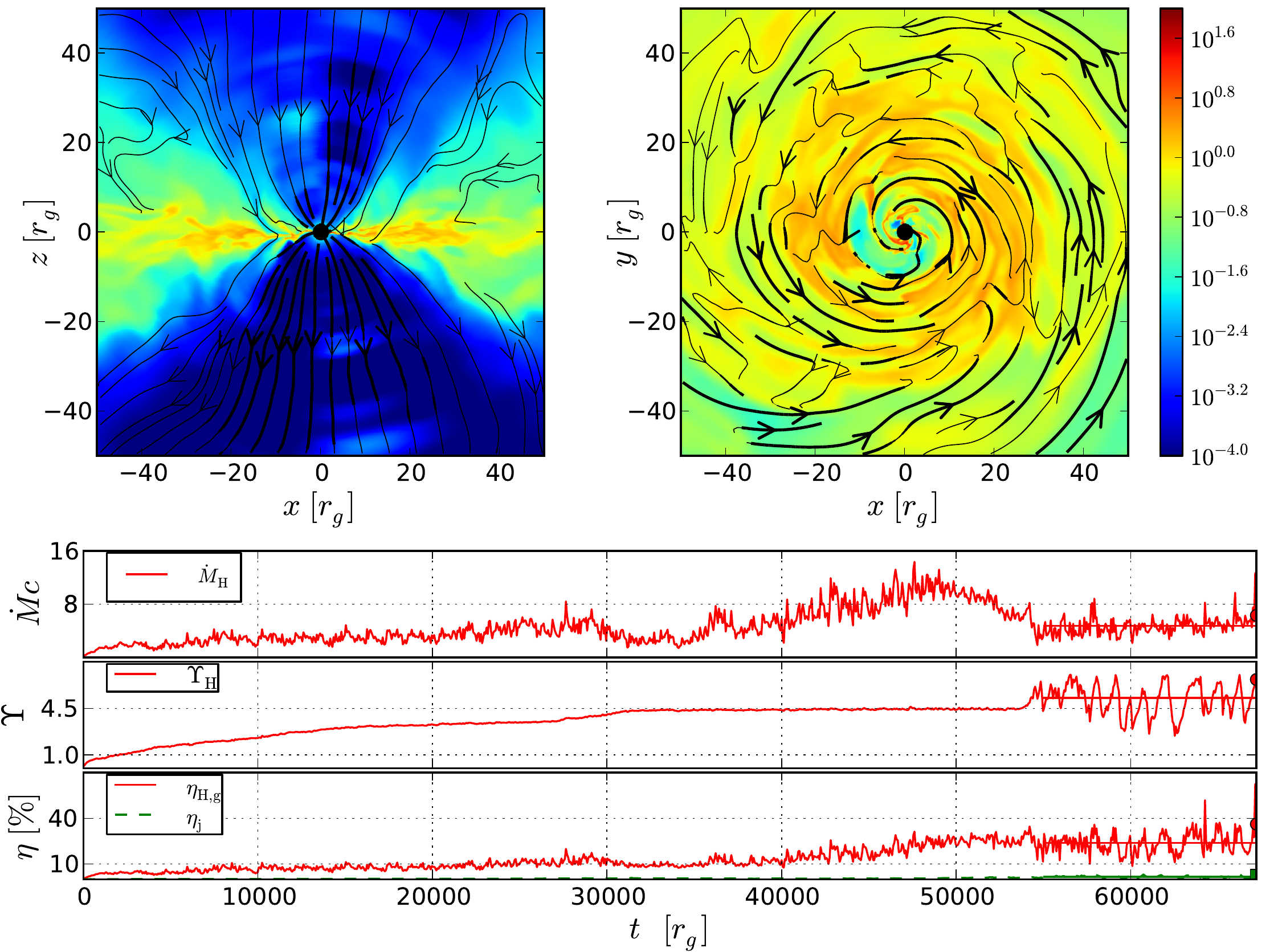}
\caption{Evolved snapshot of the MADfHR model at $t\approx
  70,000r_g/c$ showing log of rest-mass density in color (with legend)
  in both the $z-x$ plane at $y=0$ (top-left panel) and $y-x$ plane at
  $z=0$ (top-right panel).  Black lines with arrows trace field lines,
  where thicker black lines show where field is lightly mass-loaded.
  The bottom panel has 3 subpanels.  The top subpanel shows the mass
  accretion rate through the BH ($\dot{M}_{\rm H}$).  The middle
  subpanel shows the magnetic flux threading the BH ($\Upsilon_{\rm
    H}$), normalized so that order unity is a dynamically-substantial
  amount of magnetic flux.  The bottom subpanel shows the gas
  efficiency measured on the horizon ($\eta_{\rm H,g}$) and jet
  efficiency measured at $r=50r_g$ ($\eta_{\rm j}$).  In this model,
  the disk initially has a weak large-scale poloidal magnetic field,
  which is transported by the MRI and other magnetic stresses toward
  the BH.  Eventually, by $t\sim 55000r_g/c$, the magnetic flux
  saturates on the BH and cannot grow despite plentiful magnetic flux
  outside.  The MAD state that develops consists of a highly
  non-axisymmetric clumpy thin disk with a weak ($\eta_{\rm j}\sim
  1\%$) jet.}
\label{fig:MADfHR}
\end{figure*}

The initial disk is Keplerian with rest-mass density that is Gaussian
in angle with a half-height-to-radius ratio $H/R\approx 0.1$ and a
power-law radial distribution $\rho\propto r^{-0.6}$. Since this is
not an equilibrium solution near and inside the ISCO at $r_{\rm
  ISCO}$, the density is tapered as $\rho\rightarrow\rho(1-.95(r_{\rm
  ISCO}/R)^3)$ and truncated at $r_{\rm ISCO}$. The gas internal
energy density $e_{\rm gas}$ is obtained from the vertical hydrostatic
equilibrium condition of $H/R\approx \hrat \equiv c_s/v_{\rm rot}$ for
sound speed $c_s\approx\sqrt{\Gamma P_{\rm gas}/\rho}$ and rotational
speed $v_{\rm rot}$ chosen to be Keplerian with velocity $v_K=
(r/r_g)/((r/r_g)^{3/2}+a/M)$. In order to seed the MRI we use a $10\%$
random perturbation to the pressure that obeys the ideal gas law
$P_{\rm gas}=(\Gamma-1) e_{\rm gas}$.  We choose $\Gamma=4/3$ because
the disk is expected to be radiation dominated near the BH. We set an
atmosphere for the disk with $\rho=10^{-4}(r/r_g) ^{-3/2}$ and $e_{\rm
  gas}=10^{-6}(r/r_g)^{-5/2}$.

We choose an initially purely poloidal large-scale magnetic field.
With other low resolution models already performed, we know ahead of
time the magnetic flux reached on the BH and throughout the disk in
the quasi-steady MAD state, so we then tune our initial conditions to
be either nearly MAD (for MADi) or very sub-MAD (for MADf).  Net
magnetic flux on the BH and through the disk out to some radius is
conserved and only changes by accretion or expulsion through that
radius.  To reach the maximal amount of magnetic flux threading the BH
and disk, we provide a large supply, and the amount of supply ends up
not mattering because it is vast compared to what ends up on the BH or
in the disk near the BH.

The coordinate basis $\phi$-component of the vector potential is
matched to the radial gas pressure profile so that the plasma
parameter $\beta=P_{\rm gas}/P_b$ ($P_b$ as magnetic pressure), and
thus the number of critical MRI wavelengths spanning a full
scale-height of the disk, is a tunable constant for all radii. For
MADiy runs, we found it useful to transition from MAD-flux levels to a
weaker sub-MAD field beyond a transition radius, $r_{0,\rm MAD}=30$,
with a transition function $F_{\pm}=0.5\pm\frac{1}{\pi}
\arctan((r-r_{0,\rm MAD})/4)$. The vector potential is then calculated
by radially integrating
\begin{equation}\label{eqn:vecpot}
A_{\phi,r}\propto(sF_-+F_+)  \sin\theta^\nu \cdot
        \begin{cases}
        B^\theta_{0} & \mbox{if } r > r_{\rm tr} \\
        B^\theta_{tr,0} r^{(-0.6/2+3/2)/\mu} & \mbox{if } r <= r_{\rm tr}
        \end{cases}
\end{equation}
where $B^\theta_{0}$ is the field threading the disk mid-plane at time
t=0 necessary to start with $s=2$ and $\beta=200/s^2=50$ inside
$r_{0,\rm MAD}$, and with $\beta=200$ elsewhere. Thus, the disk within
$r_{0,\rm MAD}$ initially has $S_\mathrm{d}\approx0.7$, where
$S_\mathrm{d}$ is the number of MRI wavelengths within two disk scale
heights, $S_\mathrm{d}\lesssim0.5$ where the MRI is suppressed
\citep{1998RvMP...70....1B}, and MADs typically have had
$S_\mathrm{d}\lesssim0.25$ \citep{2012MNRAS.423.3083M}.  This means
the MRI is marginally suppressed in MADs.  The field threading the BH
horizon and within the ISCO matches to the disk values at a transition
radius of $r_{\rm tr}=12$. $\mu=4$ ensures the initial flux residing
within $r_{\rm tr}$ is small enough ($\approx 2.5$ times smaller than
what ends up on the BH plus out to $r_{\rm tr}$) that the BH and the
region between the horizon and $r_{\rm tr}$ can slowly build-up to a
MAD level.

To ensure a MAD can readily develop for a thin disk, our MADf model
starts with sub-MAD flux at t=0, with $\beta=200$ outside $r_{\rm
  tr}$.  The solution for $A_\phi$ is the same as the $r > r_{\rm tr}$
branch of Eq.~\ref{eqn:vecpot} but with $r_{0,\rm MAD}\rightarrow
-\infty$ and a linear taper in the field from $r_{\rm tr}$ to the ISCO
so there is no initial magnetic flux any closer.

\subsection{Numerical Grid and Density Floors}

We use similar numerical grid mapping and identical boundary
conditions as \citet{2012MNRAS.423.3083M}, except a smooth arctan
transition from exponential to hyper-exponential radial grid spacing
is used. For our high (mid) resolution grid we choose resolution $N_r
\times N_\theta \times N_\phi$ = 192x96x208 (168x64x152) with radial
grid spanning $R_{\rm in}\approx 0.75r_H$ (horizon radius $r_H$) to
$R_{\rm out}=10^4r_g$.

The $\theta$-grid spans from $0$ to $\pi$ with the mapping of
\citet{2012MNRAS.423.3083M}, but with $n_{\rm jet}=0$ used for our
entire disk since it is thin at all radii. We adjust the constants of
the mapping so that at $r/r_g=6,30,100$ we resolve the disk vertically
with {56,62,60} cells within $\pm2$H$=2\cdot0.1$R. We flare the inner
radial grid to avoid unnecessarily small limiting time-steps. The
polar boundary conditions are transmissive (see appendix in
\citealt{2012MNRAS.423.3083M}), and the $\phi$-grid is equally spaced
from 0 to $2\pi$ with periodic boundary conditions.

Apart from our direct resolution studies, we also measure convergence
quality factors for the MRI and turbulence correlations lengths. Our values demonstrate
good resolution and are reported in Section \ref{sec:results:convergence}.

\subsection{Cooling}

To keep the disk thin and radiatively efficient, we implement an
\textit{ad hoc} cooling function similar to \citet{nkh10} with
\begin{equation}
S_\nu=\left(\frac{dU}{d\tau} \right) u_\nu ,
\end{equation}
as a source term to the conservation equations, $\nabla_\mu
T^\mu_\nu=S_\nu$, for gas stress-energy tensor $T$ and 4-velocity
$u_\nu$. This assumes an isotropic comoving rate of thermal energy
loss $dU/d\tau$. $dU$ is computed so that cooling decreases internal
energy toward a target temperature set by the desired $H/R\approx 0.1$
and local density and radius.

Radiative cooling is efficient for a NT thin disk with a cooling
timescale on the order of the orbital timescale \citep{nov73}.  The NT
model assumes radiation gets out only by diffusion, while our disk is
clumpy and can be optically thin near the ISCO
\citep{2012MNRAS.424.2504Z}.  Radiative GRMHD MAD simulations are
required for an accurate cooling timescale, which could be faster than
for literal NT disks \citep{2015arXiv150802433M}.  We only try to keep
the disk thin, and we settled upon a graded cooling timescale $d\tau$ so
that
\begin{equation}
\frac{dU}{d\tau}=-\frac{\rho T_{\rm target}}{d\tau}=-\rho T_{\rm target}\cdot
        \begin{cases}
            \Omega_{\rm K}/\tau_{\rm cool} & \left(\frac{H}{\nu_{\rm Alf}}\right) > \tau_{\rm cool}/\Omega_{\rm K} \\
            \frac{\nu_{\rm Alf}}{H} & \tau_{\rm cool}/\Omega_{\rm K} > \left(\frac{H}{\nu_{\rm Alf}}\right) > 2\Delta t \\
            \frac{1}{2\Delta t}    & \left(\frac{H}{\nu_{\rm Alf}}\right)   < 2\Delta t
        \end{cases}
\end{equation}
where $\nu_{\rm Alf}$ is the Alfvenic frequency and thus $H/\nu_{\rm
  Alf}$ is the Alfven time across one disk scale height, $2 \Delta t$
is the code time-step with a prefactor 2 for stability, and
$\Omega_{\rm K}=v_{\rm K}/r$. The target temperature is from thin disk
theory and is isotropic across cylindrical radii R,
\[T_\mathrm{target}\equiv \left( \frac{0.1\cdot R}{a+r^{3/2}} \right)^2  \]
where the denominator uses spherical radius r to control the temperature
along the polar axes.
If $T_{\rm gas}=P_{\rm gas}/\rho<T_{\rm
  target}$ then $dU/d\tau=0$. To ensure there is always rapid enough
cooling, we set $\tau_{\rm cool}=0.1\cdot2\pi$ that enforces cooling
on 1/10$^{\rm th}$ the orbital timescale.

We also use a temperature ceiling that acts as another cooling term,
operating after each timestep.  We limit the temperature where
$\beta<0.1$ and $T_{\rm gas}>T_{\rm target}$.  Unfortunately these losses were
not tracked in the version of HARM used.  When we sum the radiative
losses from the cooling function during our analysis, in order to
compute the contribution lost to the temperature ceiling, we restarted
MADiHR at a late time with the ceiling turned off. The additional heat
is then captured by the cooling function which we can measure for the
radii near the BH where the temperature ceiling mattered. Accounting
for all cooling effects gives us a nearly constant radial profile for
total efficiency, and so the procedure correctly recovers how a
steady-state flow should behave.

Rest-mass density is injected whenever $2P_b/\rho>200$ in order to
ensure the code remains stable, and internal energy density is added
when $e_{\rm gas}/\rho>100$ (though this limit is never reached).

As seen in Fig.~\ref{fig:vsr} the cooling keeps the disk very close to the 
target scale height. After a very short ($<5000r_g/c$) initial period of disk evolution, the 
disk height is very stable.

\subsection{Diagnostics}
\label{ssec:diagnostics}
\newcommand{\MBH}{{M}}
\newcommand{\MBHO}{{M_i}}
\newcommand{\Mdot}{{\dot{M}}}
\newcommand{\Mdotedd}{{\dot{M}_{\rm Edd}}}

The disk's geometric half-angular thickness ($H$) per radius ($R$) is
\begin{equation}\label{eqn:thicknesseqn}
\frac{H}{R}(r,\phi) \equiv  \frac{\left(\int_\theta \rho(\theta-\theta_0)^n dA_{\theta\phi}  \right)^{1/n}}{\left(\int_\theta \rho dA_{\theta\phi} \right)^{1/n}} ,
\end{equation}
where
\begin{equation}\label{eqn:thicknesseqn2}
\theta_0(r,\phi) \equiv \frac{\pi}{2} + \frac{\left(\int_\theta \rho(\theta-\pi/2)^n dA_{\theta\phi}  \right)^{1/n}}{\left(\int_\theta \rho dA_{\theta\phi} \right)^{1/n}} ,
\end{equation}
and with surface differential $dA_{\theta\phi}$.  Typically we use $n=1$
but also compare to using other $n$.

The radiative energy and angular momentum lost are
\begin{equation}
L_\nu(r) = \int R^r_\nu(r) dA_{\theta\phi} = -\frac{1}{Dt}\int_{Dt}\int_{R_{\rm cap}}^{r}S_\nu dt dV,
\end{equation}
respectively, for $\nu=t,\phi$ for some period $Dt$ over all angles
with photon capture radius $R_{\rm cap}$ (see
\textsection\ref{sec:results}). The radiation contribution from beyond
the inflow equilibrium radius ($\sim 30r_g$ for our MADiHR model) is
conservatively estimated by using NT's radial dependence.  

The mass accretion rate, energy efficiency, and specific angular
momentum accreted are, respectively,
\begin{eqnarray}\label{Dotsmej}
\Mdot  &=&  \left|\int\rho u^r dA_{\theta\phi} \right| , \\
\eff   &=& -\frac{\int (T^r_t+\rho u^r+R^r_t) dA_{\theta\phi}}{[\Mdot]_H} , \\
\jmath &=& \frac{\int (T^r_\phi+R^r_\phi) dA_{\theta\phi}}{[\Mdot]_H}
\end{eqnarray}
where $R$ is the radiation stress-energy tensor and $[\Mdot]_H$ is the
time-averaged $\dot{M}$ on the horizon. These quantities are
calculated as totals and for the jet and wind.  The jet is defined as
where $2P_b/\rho>1$, and the wind is where $2P_b/\rho<1$ with an
outgoing flow.  The radiative efficiency and specific angular momentum
vs. radius are, respectively, $\eta_{\rm r}=-L_t(r)/[\Mdot]_H$ and
$\jmath_{\rm r}=L_\phi(r)/[\Mdot]_H$.

The dimensionless magnetic flux, that measures the strength of the
magnetic field threading the BH or disk, is
\begin{equation}\label{diag:upsilon}
\Upsilon(r) \approx 0.7\frac{\int dA_{\theta\phi} 0.5|B^r|}{\sqrt{[\Mdot]_H}} ,
\end{equation}
for radial magnetic field strength $B^r$ in Heaviside-Lorentz units,
where $\Upsilon_{\rm H}$ is $\Upsilon$ measured on the horizon and
\[\Psi(r)=\int dA_{\theta\phi} 0.5|B^r| \]is the radial absolute flux passing  
through a given radius.

Viscous theory gives a GR $\alpha$-viscosity estimate for $v_r$ of
$v_{r,\mathrm{visc}} \sim -G\alpha (H/R)^2\abs{v_\mathrm{\phi}}$, as
defined in \cite{2012MNRAS.423.3083M}, leading to a definition for the
 measure of effective $\alpha$-viscosity
\begin{equation}\label{eqn:alpha1}
\alpha_\mathrm{eff} \equiv \frac{v_r}{v_\mathrm{visc}/\alpha},
\end{equation}
whereas viscosity from local shear-stress in the disk is
\begin{equation}\label{eqn:alpha2}
\alpha_b \approx \alpha_\mathrm{mag} =-\frac{b_r b_\phi}{p_b},
\end{equation} 
where we have dropped other (e.g. Reynolds) terms from the full $\alpha$
because they are negligible compared to the magnetic term in the MAD
regions of our simulations. These $\alpha$'s (see
Fig.~\ref{fig:qsandalphas}) are obtained by separately
volume-averaging the numerator and denominator in $\theta,\phi$ for
each $r$ with a weight of $\rho$ and only including material with
$b^2/\rho_0 <1$ in order to focus the measurement on the disk material
and not the lower density corona. The quantity $\alpha_b$ measures the
magnetic stress within the dense regions, while the
$\alpha_{b,\mathrm{eff}}$ measures the actual effective $\alpha$ based
upon the radial velocity of the dense regions.  For example, very
little magnetic flux may thread the dense regions, but external
magnetic torques may push and drive angular momentum transport and
inflow of the dense regions.  $\alpha_{b,\mathrm{eff}}$ is only accurate
far outside $r_\mathrm{ISCO}$ because of the $G$ term.

%% file: msdiscuss.tex
\section{Results and Discussion}
\label{sec:results}

With our fully global 3D GRMHD simulations we can demonstrate whether
the non-linear MRI and other MHD physics in thin accretion flows is
able to transport large-scale magnetic flux towards the BH and
build-up the magnetic field to the MAD state. We first consider a disk
with a weak magnetic field using our MADiHR setup to see if
magnetic flux transports inward.  Then, having confirmed thin disks
transport magnetic flux efficiently to the BH and inner-radial disk,
we use our second setup MADiHR to study a MAD where magnetic
saturation is reached to large radii.  This two-step procedure is
necessary due to the extreme computational resources that would
otherwise be necessary to build-up large-scale flux to the MAD state
out to large radii.

\subsection{Demonstration of flux accumulation to MAD state}
\label{sec:buildup}

First, we consider a thin radiatively efficient disk threaded by weak
large-scale magnetic field.  This model, MADfHR/MR, started in the
sub-MAD state with $\beta\approx 200$ (a field strength that is $\sim
7$ times weaker than when the MRI is suppressed and $\sim 40$ times
weaker than the final MAD state) so that the MRI operates over long
times.  In MADfHR the MRI leads to accretion of magnetic flux inward,
such that over $t\approx70,000r_g/c$ the BH and disk become MAD out to
$r\sim 15r_g$. Fig.~\ref{fig:MADfHR} shows a steady accumulation of
flux during the first $\sim30000r_g/c$. The level of flux then remains
constant on the horizon for another $\sim20000r_g/c$ due to an excess
of rest-mass accumulating in the inner disk that slowly accretes. This
denser torus forms as the magnetic flux threading the inner section of
the disk accretes onto the BH during the early evolution and
temporarily starves some disk material of strong field. The local
viscosity is then suppressed until magnetic flux from further out in
the disk has time to transport inward. Once this initial dense torus
accretes, the disk enters a MAD regime of accretion.

In the much longer run, MADfMR becomes MAD out to a radius of $r\sim
18r_g$ by the time $t\approx183,000r_g/c$.  Plenty more magnetic flux
exists at larger radii, but it would presumably contribute to the MAD
region of the disk at later times.  The physics of magnetic flux
accumulation involves the weak field MRI driving accretion, magnetic
braking via the wind transporting material and magnetic flux inward,
and may involve the field penetrating coronal-like material in the
disk (e.g., see \citealt{2008ApJ...677.1221R,bhk09}).  At early times,
magnetic flux at higher latitudes transport more readily than those in
the equator, but eventually magnetic flux threading the disk is also
transported to the region close to the BH.  Whatever is the physics
behind the saturation of magnetic flux, it has reached a maximum far
below the supply, yet far beyond the initial value. This demonstrates
that large-scale magnetic flux can readily accumulate even with
$H/R\approx 0.1$ and radiative cooling, and that it reaches a
saturated value like seen in prior MAD simulations.

\begin{figure}
\includegraphics[width=3.15in,clip]{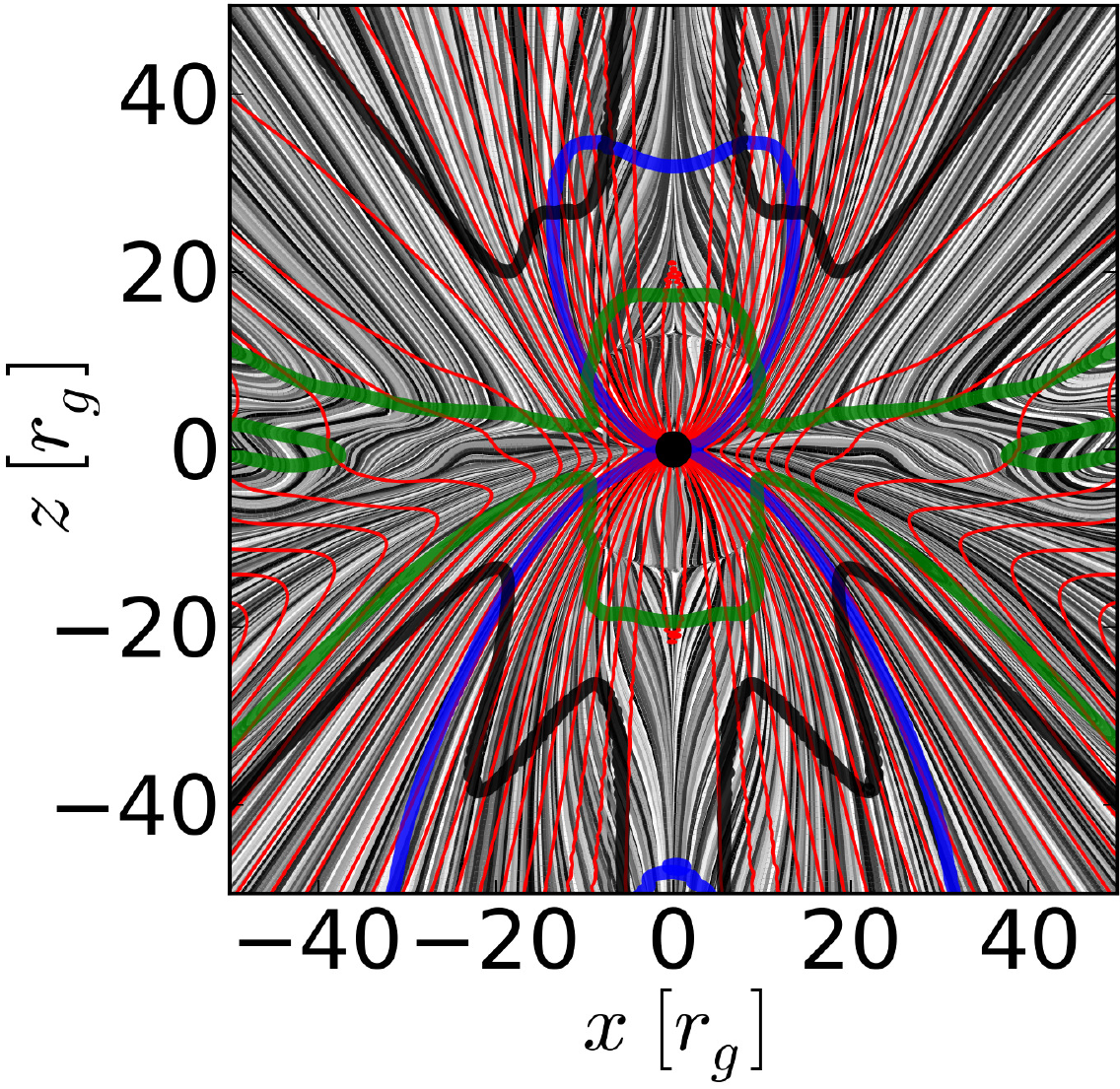}
\caption{Time and $\phi$-averaged quantities are plotted for
  MADfHR. Rest mass gas flow is traced by black streamlines. The
  colored (green, black, and blue) thick lines correspond to
  time-averages of quantities $Y=\{u^r,u_t,b^2/\rho_0\}$,
  respectively, at values $[Y]_t=\{0,-1,1\}$.  While near the BH the
  flow has $[b^2/\rho_0]_t\gtrsim1$ as averaged directly, the dense
  inflow has $[b^2/\rho_0]_t\lesssim1$ at all other radii.
  Streamlines nearest the polar axes are calculated using velocity
  structure five cells away from the axis for better visualization of
  the qualitative geometry of the jet flow.  The flow exhibits
  equatorial asymmetry over many inflow times as found in thicker MAD
  simulations. The red magnetic field line contours show how the
  magnetic flux threads the black hole and inner
  disk.} \label{fig:avgMADfHR}
\end{figure}

Fig.~\ref{fig:avgMADfHR} shows the time and $\phi$-averaged rest-mass
velocity streamlines as well as several key boundaries associated with
the accretion flow.  This includes showing the jet boundary and the
flow stagnation surface (where material has zero radial velocity,
moves outward at higher latitudes, and moves inward at lower
latitudes). Most importantly, the plot indicates the
time-$\phi$-averaged structure of the magnetic field once the inner
disk and BH have reached magnetic flux saturation. As with the thicker
MAD model plotted similarly in Fig. 6 of \cite{2012MNRAS.423.3083M},
the time-$\phi$-averaged steady-state field structure, as well as the
evolution we see in movies, demonstrate ordered poloidal magnetic flux
is easily transported through the disk from large radii.

\begin{figure*}
\centering
\includegraphics[width=6.3in,clip]{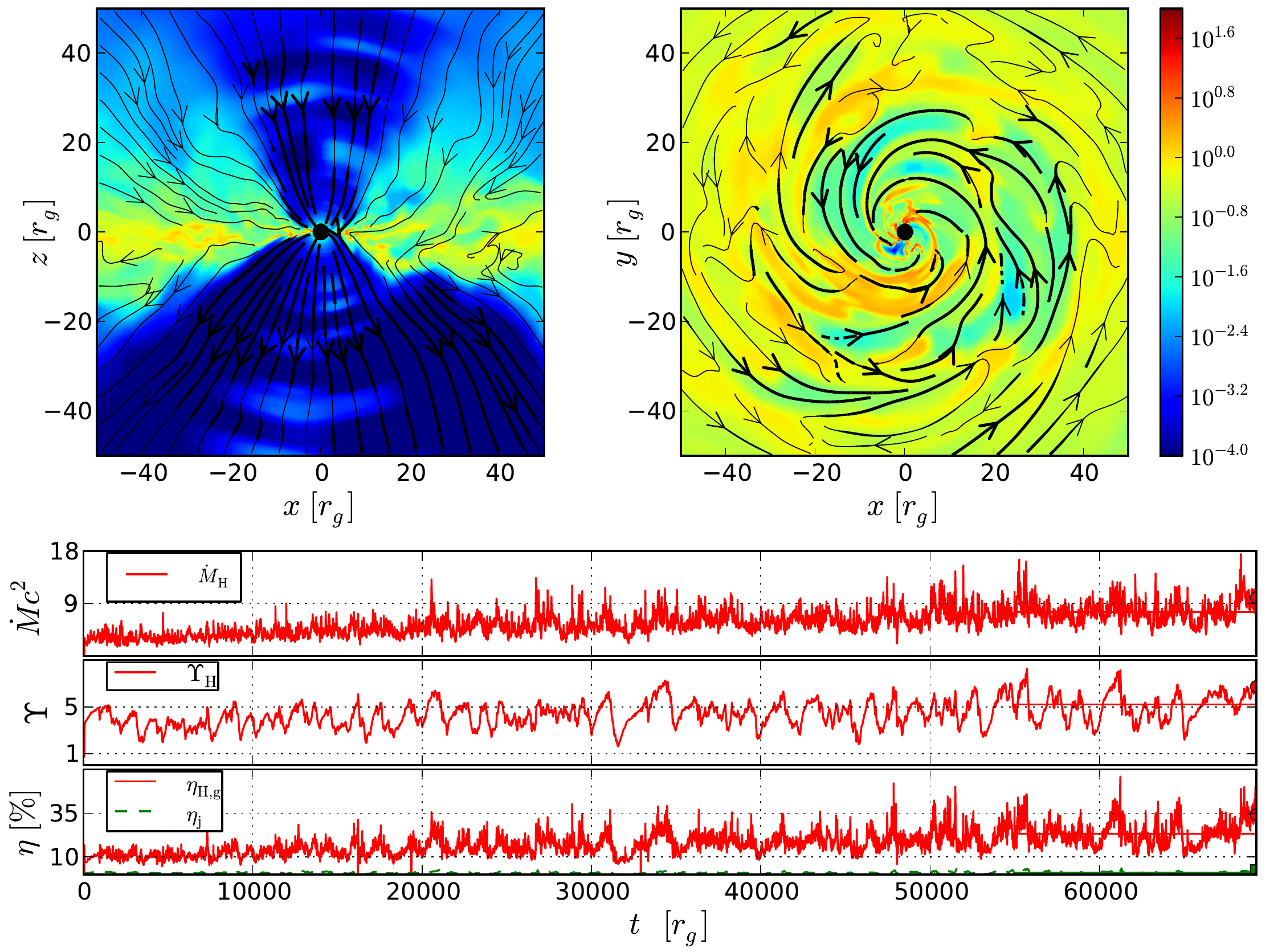}
\caption{Evolved snapshot of the MADiHR model at $t\approx
  70,000r_g/c$, with identical layout to Fig.~\ref{fig:MADfHR}. The
  MAD state is highly non-axisymmetric and clumpy with a weak
  ($\eta_{\rm j}\sim 1\%$) jet. The rest-mass accretion rate and
  magnetic flux on the horizon reach the MAD state quickly, which was
  as desired so that we can study a quasi-steady MAD out to large
  radii over long time periods.  The dimensionless magnetic flux on
  the horizon is in agreement with our MADfHR model, showing a disk
  initial condition with a relatively stronger poloidal field strength
  still reaches the same MAD state near the BH.}
\label{fig:evolved}
\end{figure*}

\subsection{Long-term evolution of MAD state}
\label{sec:initmad}

\begin{figure}
\includegraphics[width=3.35in,clip]{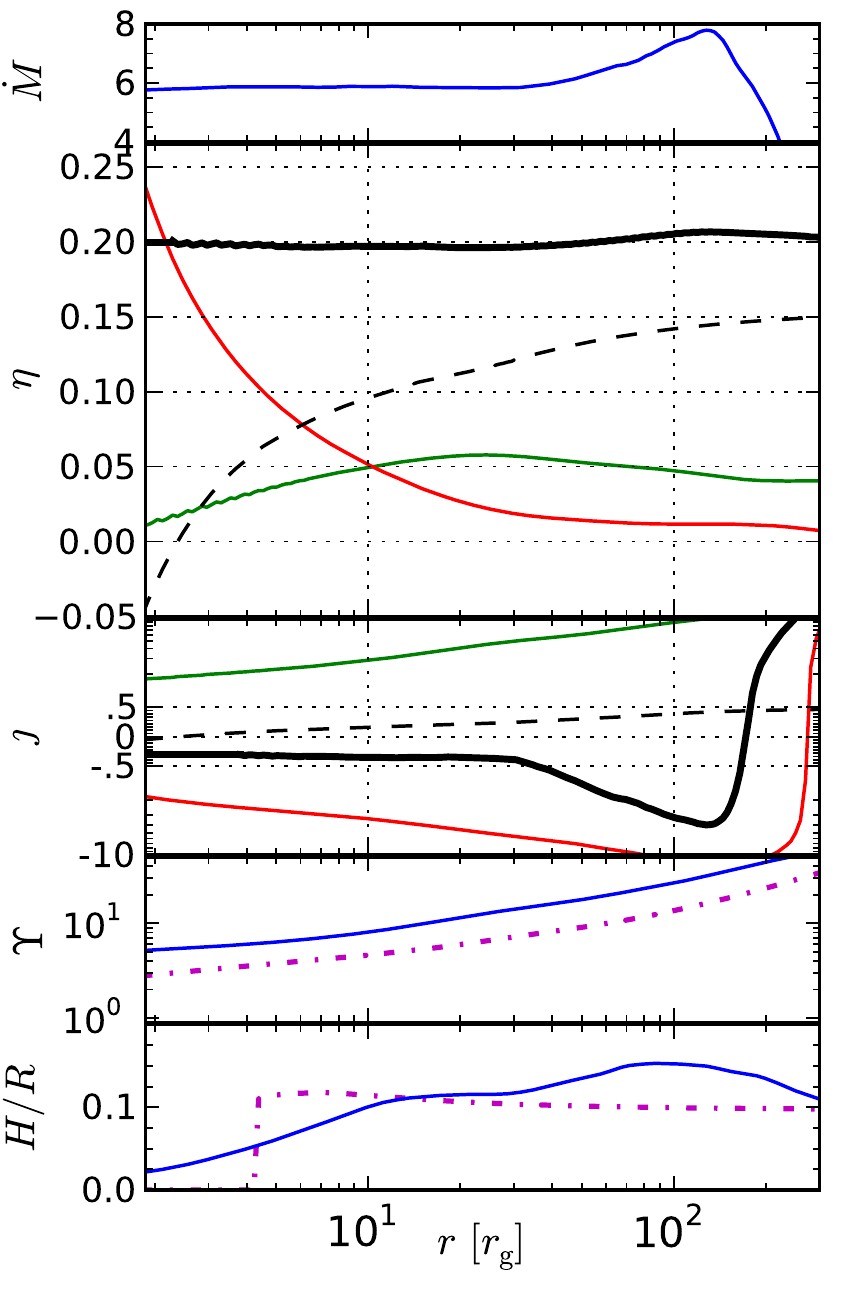}
\caption{Time-averaged quantities as a function of radius. From top to
  bottom, panels are: Total mass accretion rate ($\Mdot$, blue line),
  energy efficiency ($\eta$), specific angular momentum ($j$),
  magnetic flux ($\Upsilon$), and disk thickness ($H/R$, n=1 in
  Eq.~\ref{eqn:thicknesseqn}, blue line). 
  Plots for $\eta$ and $j$ show total gas+radiative (black
  solid), electromagnetic (green), matter (red), and radiation (black
  dashed) components.  Remaining plots show their initial value
  (magenta dot-dash). The plot of $\Upsilon$ shows that for the total
  BH+disk contribution (blue). The constancy of $\dot{M}$, $\eta$, $j$
  show where the flow has achieved inflow equilibrium.  $\Upsilon$
  shows the MAD to non-MAD transition at $r\sim 35r_g$,
  while throughout the disk $H/R\approx 0.1$ (bound gas), as
  maintained by cooling.  The primary result of this plot is the
  efficiency panel showing the radiative contribution of $\eta_{\rm
    r}\sim 15\%$ with most of emission beyond the standard NT disk
  model coming from quite close to the photon orbit.} \label{fig:vsr}
\end{figure}

\begin{figure}
\centering
\includegraphics[width=3.35in,clip]{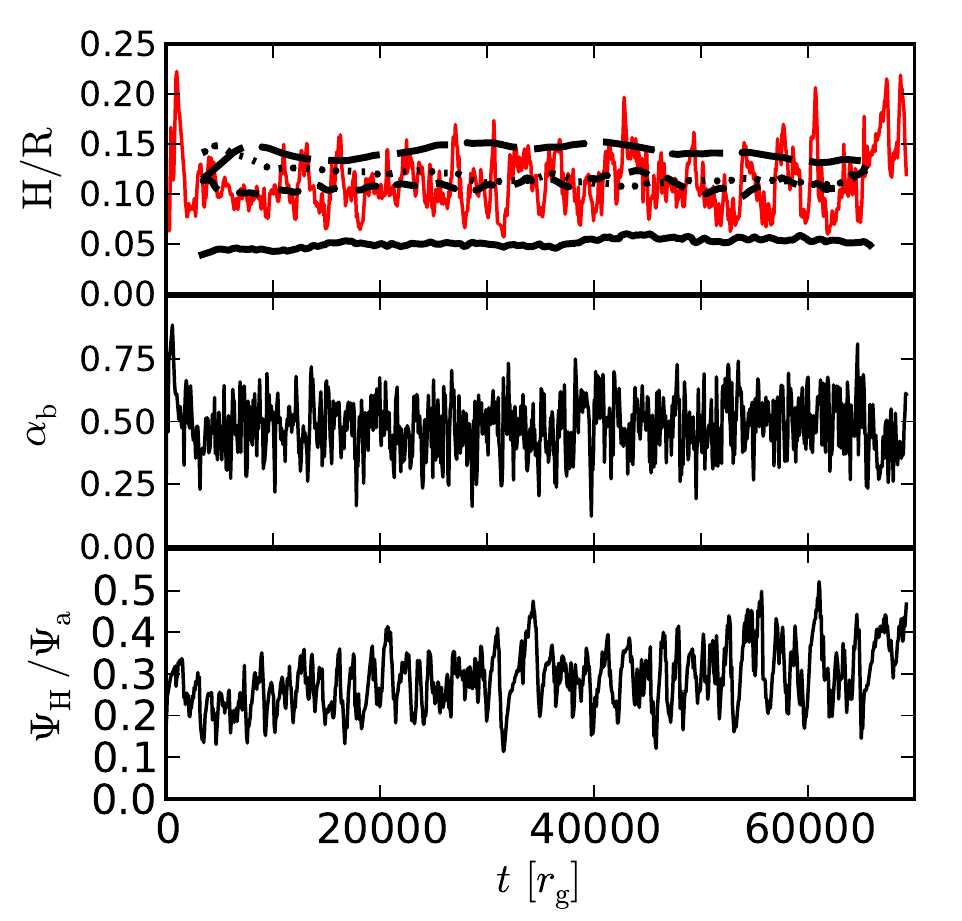}
\caption{Shows temporal evolution of a few quantities for our MADiHR
  model. The disk scale height $H/R$ is shown in the top panel. We
  show the temporally-smoothed $H/R$ as plotted for radii
  $R_\mathrm{ISCO}$ (solid-black), $15r_g$ (short-dashed), $30r_g$
  (dots), and $60r_g$ (long-dashed). The red line is the un-smoothed
  evolution at $15r_g$ to demonstrate the level of variability.  The
  $H/R$ value is regulated by cooling to be $H/R\sim 0.1$, which is
  what is achieved for radii just outside the ISCO and out to the
  inflow equilibrium radius of $r\sim 35r_g$. The middle panel shows
  the $\alpha_b$ viscosity parameter spatially-averaged over the disk
  at a radius $10r_g$.  The relatively high $\alpha_b\sim 0.5$ is
  indicative of MADs, implying efficient transport of mass and angular
  momentum. The third panel shows the ratio of the total magnetic flux
  threading the BH to the total magnetic flux available at the start
  of the simulation inside the radius out to which the simulation
  ultimately reaches at least one inflow time.  This shows that the
  horizon contains much more magnetic flux than is available, so the
  quasi-steady state has a horizon-threading flux that has reached
  saturation.}
\label{fig:MADiHRvst}
\end{figure}

In order to study the thin MAD out to large radii to ensure an
accurate quasi-steady radiative efficiency, we next considered the
disk having an initially nearly MAD-level of magnetic field.
Fig. \ref{fig:init} and Fig. \ref{fig:evolved} show the initial and
evolved quasi-equilibrium state of our high-resolution model that is
initially nearly MAD inside $r=30r_g$.  As the accretion flow evolves
from its initial state, rotation drives outward angular momentum
transport by turbulent magnetic stresses via the MRI as well as by a
wind. Mass, energy, angular momentum, and magnetic flux are accreted
and ejected, and by $\sim5000r_g/c$ magnetic stresses balance incoming
gas forces so the accretion is MAD out to $r\sim30r_g$. By the end of
the simulation, $t\sim70000r_g/c$ for MADiHR, the MAD extends to $r\sim35r_g$.

\begin{figure}
\includegraphics[width=3.35in,clip]{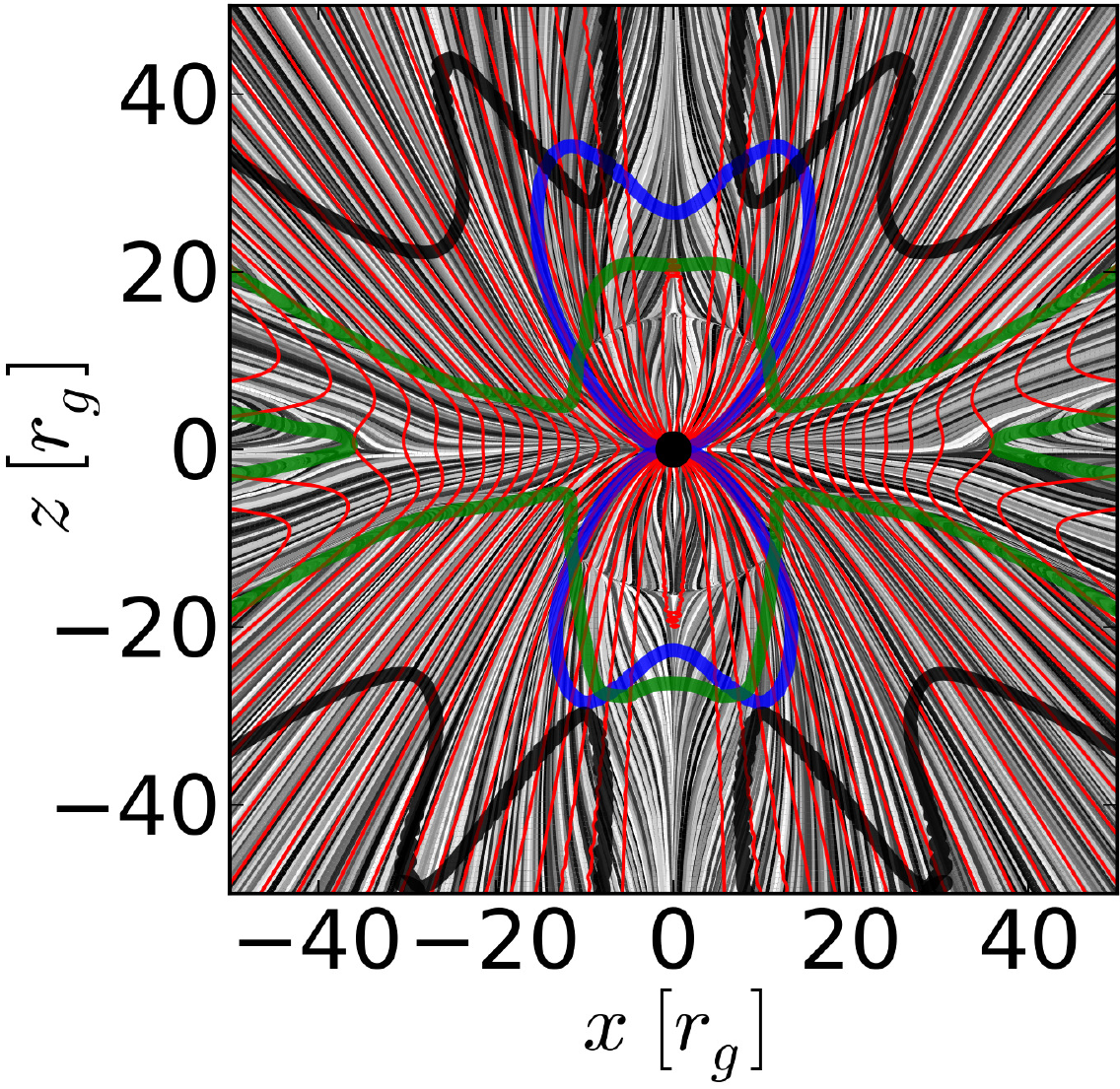}
\caption{Time and $\phi$-averaged quantities are plotted for
  MADiHR. Identical layout as Fig.~\ref{fig:avgMADfHR}. The disk is
  magnetically arrested out to a transition radius of
  $\sim35r_\mathrm{g}$, which leads to a short inflow time-scale
  compared to the MADfHR disk at those radii, so the time-averaged
  velocity fields and other disk structure is smoother and more
  symmetric. } \label{fig:avgMADHRNT}
\end{figure}

The evolved state is evaluated using the period $t=[30000,
  70000]r_g/c$ for time-averaging quantities.  A slightly longer
period $t=[10000,70000]r_g/c$ is used for the time-averaged functions
of radius in Fig.~\ref{fig:vsr} to minimize the effects of long-term
trends. The time-$\phi$-averaged structure is show in
Fig.~\ref{fig:avgMADHRNT} and shows qualitatively similar behavior to
MADfHR.

The gas+radiative efficiency
$\eff\sim 20\%$, and specific angular momentum $\jmath\approx -0.4$
are constant within the MAD/non-MAD transition $r\sim 35r_g$,
indicating a quasi-steady MAD state has been reached out to a
relatively large radius.

The disk thickness at $r\sim 20r_g$ is regulated by cooling to be
$H/R\approx 0.11$ for n=1 in Eq.~\ref{eqn:thicknesseqn},
$H/R\approx0.14$ for $n=2$, density-weighted normalized $H/R\approx
0.13$ \citep{2012MNRAS.420..684P}, and thermal thickness of
$\hrat\equiv \arctan{(c_s/v_{\rm rot})}\approx 0.13$.  An accurate
measurement of $H/R$ is complicated by the fact that magnetic
compression leads to a geometric $H/R\approx 0.05$ near the BH, mass
is launched into a broad wind, and vertical disk fluctuations are the
same order as $H/R$.  For the $\phi$-averaged density profile and
including only bound material (i.e. $-u_{t}(\rho+e_{\rm gas}+P_{\rm g}
+ 2P_b)/\rho <1$), valid when comparing with NT, we get $H/R\approx
0.10$, which is our target value we desired for the disk proper.
Fig.~\ref{fig:MADiHRvst} shows the time evolution of $H/R$ at four
radii for $n=1$.

Our thin MAD achieves a magnetic flux on the BH of $\Upsilon_{\rm
  H}\approx 5$.  The stability of $\Upsilon_{\rm H}$ with time and the
presence of plentiful magnetic flux remaining in the disk (see
Fig.~\ref{fig:MADiHRvst}, lower panel) indicate that the disk has
achieved a balance between magnetic forces and gas forces with a
strong magnetic field ($\Upsilon\gg 1$). Thus one expects maximum
magnetically-induced deviations from the standard NT thin disk
solution.

\begin{figure*}
\centering
\includegraphics[width=6.5in,clip]{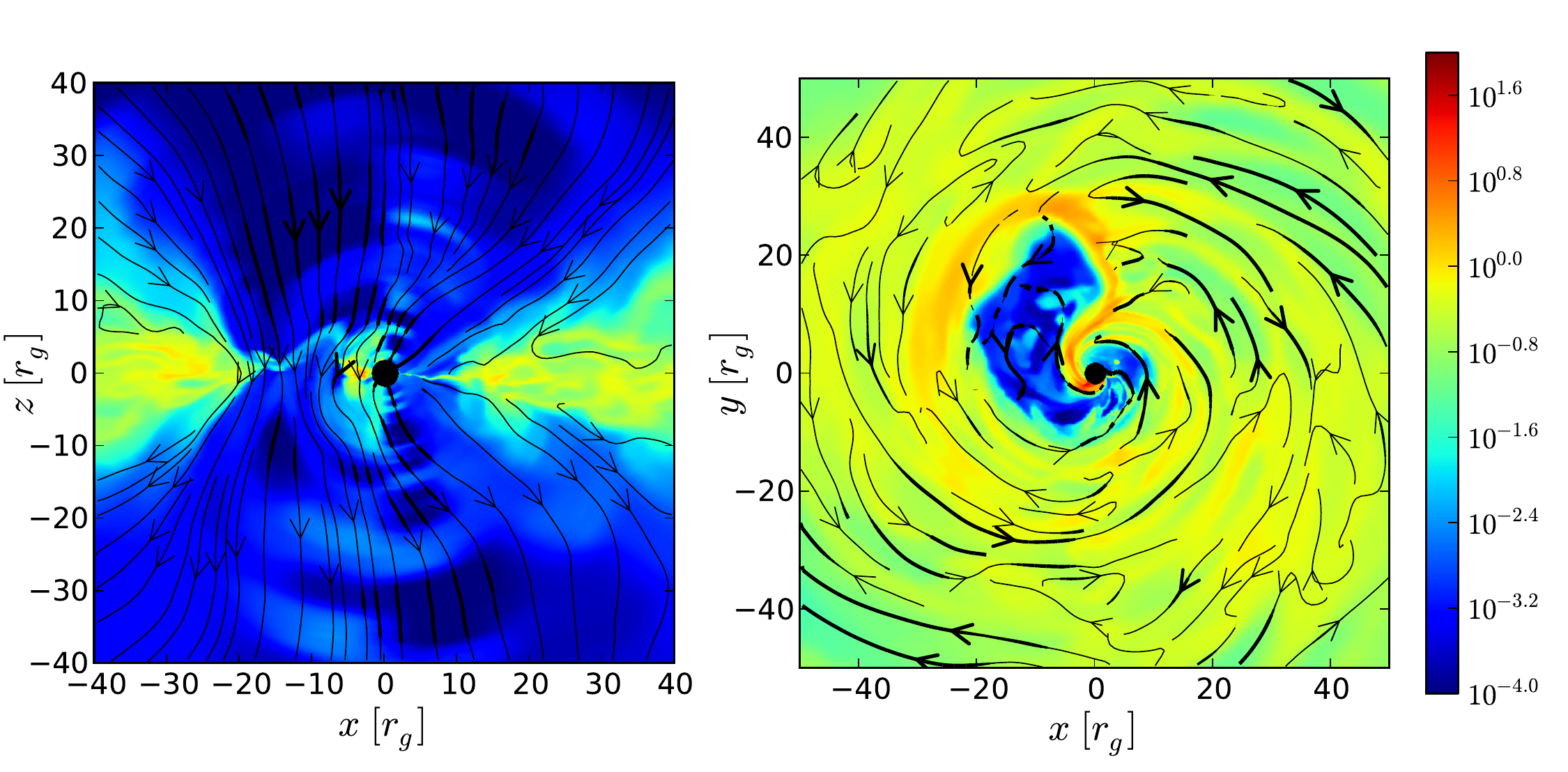}
\caption{Plotted as in the top frames of Fig.~\ref{fig:evolved} is a 
snapshot of MADiHR at $t=31568r_g/c$ when the disk experiences an extreme eruption of 
magnetic flux from the BH horizon. A large magnetic flux tube creates a low-density cavity 
in the disk before it is partially blended with disk material 
by the magnetic Rayleigh-Taylor instability. Such prominences occur frequently in the MAD state.}
\label{fig:evolvedExtreme}
\end{figure*}

Large temporal deviations are driven by magnetic flux repeatedly
building up on the BH horizon to a level the disk cannot sustain.  As
soon as a weak point in the density exists caused by a
non-axisymmetric mode, then there is a sudden re-emergence of magnetic
flux back into the disk, a process that creates large low-density
bubbles in the form of a magnetic prominence.  The magnetic flux that
is ejected is then processed by turbulence by the outer parts of the
disk until being accreted again at which point the process
repeats. Fig.~\ref{fig:evolvedExtreme} shows one of the most extreme
magnetic prominences occurring in the MADiHR simulation.

\subsection{Radiative Efficiency}

We find that nearly as much radiation is released near and inside the
ISCO as in the entire rest of the disk, so we calculate the radiative
efficiency assuming 100\% photon capture within a choice among three radii: 
the horizon, the photon orbit radius $r\approx2.4$, and the critical 
impact parameter radius,
$r\approx4.1$, for which geodesics with any smaller impact parameter intersect
the horizon.  An estimate of photon capture
using ray-tracing from a thin disk (see \citealt{2006astro.ph..5295B})
gives as much as $30\%$ escape fraction from the photon orbit for
equivalent BH spin, and more than $70\%$ for $r\approx4.1$. This
suggests using the photon orbit may give the best, simplest
approximation.

The radiative efficiency is $\eta_{\rm r}\sim 15\%$ for MADiHR
($<0.5\%$ lower for MADiMR) when including only radiation from outside
the photon orbit.  This is about twice larger than the $\eta_{\rm
  r}\approx 8\%$ efficiency in standard NT thin disk theory for BH
spin $a/M\approx 0.5$. The impact parameter estimate is an extreme,
albeit unphysical, limit giving efficiency $\eta_{\rm r}\sim 9.4\%$.
Also unphysical, including all radiation released outside the horizon
gives $\eta_{\rm r}\sim 19\%$. Unbound gas only contributes $\sim 1\%$
to $\eta_{\rm r}$, a smaller relative contribution compared to prior
simulations \citep{2010MNRAS.408..752P,2012MNRAS.424.2504Z}.

Thus, our thin MAD with $H/R\approx 0.05$--$0.1$ and a radiative efficiency
of $\eta_{\rm r}\approx 15\%$, demonstrates an $\approx 80\%$ deviation from
NT.  Prior GRMHD simulations of thin disks have had $H/R\approx 0.1$
with $6\%$ deviations \citep{2009ApJ...692..411N}, $H/R\approx 0.06$
with $6\%$ to $10\%$ deviations \citep{2011ApJ...743..115N}, and
$H/R\approx 0.07$--$0.13$ with $4\%$ to $5\%$ deviations
\citep{2010MNRAS.408..752P}.  So despite having similar $H/R$, our
thin MAD has vastly higher deviations from NT.

\subsection{Jet Efficiency}
\label{sec:results:jet}

Some of the horizon magnetic flux reaches into the polar regions
leading to a jet efficiency of $\eta_{\rm j} \approx 1\%$ at $r=50r_g$
for both MADiHR and MADfHR.  A caveat is that the jet within
$b^2/\rho_0>1$ at $r=50r_g$ is only resolved by about 7 grid cells.
So we check the jet efficiency and resolution on the horizon, where
the jet is resolved with more than 12 grid points per hemisphere and
still has a jet power of $\eta\approx 1\%$.  These horizon jet grid
cells are sufficient resolution to capture the force-free behavior of
the jet near the polar axes, according to prior resolution studies
\citep{2004ApJ...611..977M}.  So the jet efficiency measurement is
robust to resolution changes, slight model changes, and what radius it
is measured at.


The BH spin and disk rotation drive a magnetized wind with efficiency
$\eta_{\rm w}\approx 4\%$, though at least an additional $1\%$ wind
efficiency may be lost to radiation via the very efficient cooling we
use as well as lost to the comparatively low resolution in the jet and
highly magnetized wind regions.

The jet and wind contribute only a bit to the total $\eff\approx
20\%$.  However, the jet and wind carry nearly as much angular
momentum outward as the dense Keplerian disk carries inward.  This
results in the BH spin-up parameter $s= -\jmath-2(a/M)(1-\eff)\approx
-0.4$, i.e. somewhat spinning down \citep{2012MNRAS.423.3083M}.

Further, the jet efficiency as a function of $H/R$ is constrained by our thin
MAD result. Magnetic compression leads to varying geometric $H/R$
vs. radius, so the more constant thermal thickness $\hrat\equiv
\arctan{(c_s/v_{\rm rot})}$ is used \citep{2012MNRAS.423.3083M}.  Our
thin MAD models have $\hrat\approx 0.13$ and older MADs had
$\hrat\approx 0.5$ (e.g. with geometric $H/R\approx 0.3$ at $r=20r_g$)
or $\hrat\approx 1.5$.

Prior thick MAD simulations found that the saturated magnetic flux on
the BH given by $\Upsilon$ is roughly constant vs. BH spin.  That data
showed a slight tendency to peak at $a/M\sim 0.5$, but this is
probably intrinsic to the initial disk conditions or due to intrinsic
errors that suggest $a/M\sim 0$ could be the peak as well
\citep{2012MNRAS.423.3083M,2012JPhCS.372a2040T}.  The thick MAD
simulations of moderate thickness show a clear asymmetry in the
$\Upsilon$ vs. spin for positive vs. negative spin that is consistent
with the asymmetry in jet efficiency vs. spin.  So we do not try to
fit the peak in $\Upsilon$ being at $a/M\sim 0.5$, but we capture the
asymmetry.  We also capture the fact that this spin asymmetry is seen
to go away as the disc becomes even thicker.

Combining the different thick MAD models with our new thin MAD model,
a rough fit of the jet power is given by
\begin{equation}\label{etajfit}
\eta_{\rm j}\sim 400\%\omega_H^2 \left(1 + \frac{0.3\omega_H}{1+2\hrat^4}\right)^2 \hrat^2 ,
\end{equation}
where $\omega_H\equiv a/r_{\rm H}$, which assumes steady activation of
a \citet{bz77} jet.  The $\eta_{\rm j}\propto \hrat^2$ scaling is
required (compared to $\propto \hrat$) to reach our thin MAD with
$\eta_{\rm j}\approx 1\%$.  That $\eta_{\rm j}\propto (H/R)^2$ is
consistent with S7.1 of \citet{2010MNRAS.408..752P} who analytically
found $\Upsilon\propto (H/R)$, and with \citet{2001ApJ...548L...9M} who
argued that $\eta_{\rm j}\propto (H/R)^2$.

This formula is roughly consistent with what is expected based upon
the BZ formula $\eta_{\rm j}\propto \Upsilon^2 \Omega_{\rm H}^2$.
Models with $\hrat\approx 0.5$ and $a/M=0.5$ gave $\eta_{\rm j}\sim
20\%$ and $\Upsilon\approx 13$.  Our thin MAD has $\Upsilon\approx 5$,
so the BZ formula suggests we should get $\eta_{\rm j}\approx 3\%$
while we got $\eta_{\rm j}\approx 1\%$.  Why the factor of three
difference?

First, $\Upsilon$ measured on the BH is not relevant to the BZ jet.
One cannot use $\Upsilon_{\rm H}$ with $\eta_{\rm j}\propto
\Upsilon_{\rm H}^2$, because in general (and as found in thicker MAD
models) magnetic flux in the jet is a fraction of $\Upsilon_{\rm H}$
and the jet only partially covers the rotating BH's surface.  Instead,
the magnetic flux passing through the jet is what is relevant to the
jet power.  The $\hrat\approx 0.5$ model had $\Upsilon_{\rm j}\approx
10$, while our model has $\Upsilon_{\rm j}\approx 4$.  These are
significantly different from the magnetic flux over the entire
horizon.  This still implies we should have $\eta_{\rm j}\approx 3\%$
for consistency with the BZ expectation.

Second, the BZ formula fails to account for the disk covering a
significant part of the potential power output
\citep{2005ApJ...630L...5M}.  The disk at the equator gets in the way,
even for MADs (especially at lower spin when less magnetic compression
occurs).  However, most of the BZ power would be at the equator where
the Poynting flux peaks.  The disk chops-off this peaked part of the
Poynting flux for the jet, giving some of it to the wind.  Some of
that BZ power is never produced in the first place due to the heavy
mass-loading by the disc. A better model of that disk part that
connects to the horizon is the \citet{1999ApJ...522L..57G} model of a
thin magnetized disc.  In that case, $\Upsilon$ locally in the disk is
much less than over the entire horizon, so the
\citet{1999ApJ...522L..57G} model predicts much less than BZ for
extraction of energy.  One cannot just fit $\eta_{\rm j}$
vs. $\Upsilon_{\rm H}$ and expect it to match BZ.

Third, we have defined the jet arbitrarily as where $b^2/\rho_0>1$.
The wind is defined as the rest of the non-radiative flux that adds up
to the total efficiency.  For our thin MAD, this is 4\%
(i.e. 20\%-1\%-15\%).  The wind efficiency is partially directly due
to the disk rotation, but some of the wind efficiency derives from
magnetic flux threading the BH that passes into the disk leading to a
BH-spin-driven wind.  Some of this BH-spin-driven wind absorbs what
would have otherwise (without a disk) become the jet and can account
for another missing 2\%, giving a total BZ-like efficiency matching
expectations.  If this were the sole effect in comparison to the
thicker MAD models, this would suggest that thin MADs may have more
powerful winds per unit jet power relative to thick MADs.

Fourth, the jet efficiency for the $\hrat\approx 0.5$ $a/M=0.5$ model
is a peak among all the otherwise identical models with different BH
spins.  Given how much the results vary for slight changes in the
model setup (like initial magnetic field strength), the $\eta_{\rm
  j}\approx 20\%$ could easily have been $\eta_{\rm j}\approx 10\%$
within systematic errors.  Then there would be no discrepancy between
our formula and the BZ approximation.

So our Eq.~(\ref{etajfit}) is not supposed to be derived from BZ
because the BZ-driven process creates both a ``jet'' and a ``wind'',
while we have defined the jet by that part which would be capable of
becoming relativistic.  So in that sense, our formula is useful in
application to real jets instead of a test of whether the theoretical
BZ extraction efficiency applies to the simulation data.  However,
only direct observational comparisons without ad hoc definitions will
test the validity of these simulations being applicable to real
systems.  The fitting formula is just a useful guide.

\subsection{Resolution}
\label{sec:results:convergence}

\begin{figure}
\includegraphics[width=3.15in,clip]{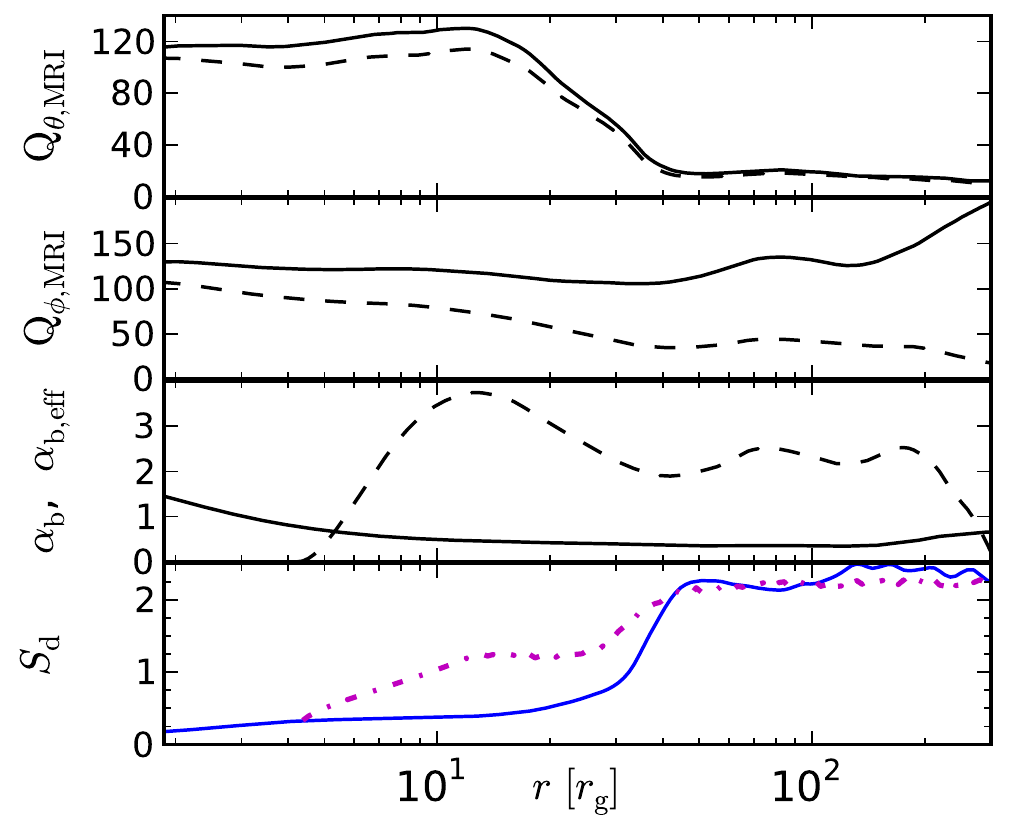}
\caption{MRI quality factors indicating time-$\phi$ averaged
  resolution, viscosity of the disk, and vertical MRI
    half-wavelengths per disk scale height ($S_\mathrm{d}$, blue
    line) are plotted as a function of
  radius. Q factors indicated the number of grid cells per critical
  MRI wavelength, averaged with weight $\sqrt{\rho b^2}$ (solid lines)
  and $\rho$ (short dashed) across the disk where $b^2/\rho<0.5$. The
  third panel shows the viscosity parameter
  \textsc{$\alpha_\mathrm{b}$} (solid line) and the effective
  viscosity calculated via the mass accretion rate (short dashed). The
  magnetic evolution of the disk is well resolved and leads to a high
  viscosity as measured by local stresses, $\alpha_\mathrm{b}$, and
  the rate of mass inflow
  $\alpha_\mathrm{b,eff}$. The magenta dot-dash line represents the 
  value of $S_\mathrm{d}$ at t=0 and the time-average is blue-solid. } \label{fig:qsandalphas}
\end{figure}

We test the resolution of our simulations using convergence quality
factors for the MRI ($Q_{\theta\phi,\rm MRI}$; number of grid cells
per MRI wavelength in each direction) and turbulence ($Q_{
  r\theta\phi,corr}$; number of cells per turbulence correlation
length in each direction)
\citep{sdgn11,hgk11,2012MNRAS.423.3083M,2013ApJ...772..102H,2013Sci...339...49M,2014MNRAS.441.3177M}. For
our high resolution runs, MADiHR (MADfHR), at $r={12,30}r_g$ centered
on the equatorial flow, we have $Q_{\theta,\rm MRI}={160,55}$
($Q_{\theta,\rm MRI}={88,18}$) with a weighting of $\sqrt{\rho b^2}$
and we have $Q_{\phi,\rm MRI}={120,100}$ ($Q_{\phi,\rm MRI}={79,77}$)
with a weight of $\rho$.  Weighted averages are computed for the
numerator and denominator separately before taking the ratio to get
$Q$.  The purely density-weighted values represent a stricter
constraint on whether we have sufficient resolution (because MADs have
strong fields mostly outside the dense region, and the external
stresses are important to whether transport occurs), and these are
plotted as a function of radius in
Fig.~\ref{fig:qsandalphas}. Resolution of rest-mass and magnetic energy densities
are evaluated at four radii $r=\{r_H/r_g,4,8,30\}r_g$. For MADiHR,
rest-mass correlation values are, at each radius respectively,
$Q_{r,corr}\sim6,10,14,14$, $Q_{\theta,corr}\sim40,35,31,34$, and
$Q_{\phi,corr}\sim6,7,8,11$.  Correlation values for magnetic energy
density are, similarly, $Q_{r,corr}\sim6,12,17,15$,
$Q_{\theta,corr}\sim34,33,29,36$, and $Q_{\phi,corr}\sim8,9,12,16$.

For MADfHR, calculated for the disk once the BH has reached the MAD
state, rest-mass correlation values are, at each radius respectively,
$Q_{r,corr}\sim6,10,14,14$, $Q_{\theta,corr}\sim3,29,28,32$, and
$Q_{\phi,corr}\sim7,9,11,13$.  Correlation values for magnetic energy
density are, similarly, $Q_{r,corr}\sim6,12,16,15$,
$Q_{\theta,corr}\sim5,28,30,31$, and $Q_{\phi,corr}\sim7,9,12,14$.

These MRI and turbulent quality factors are high enough to ensure the
reported simulations are resolving all the necessary turbulent physics
and are likely converged.  We also directly convergence tested our
results by performing simulations of each setup with three
resolutions. The qualitative structure of the disk in the MR and HR
simulations are the same, but the lower resolution in MR allowed us to
run these simulations for significantly longer. The nature of magnetic
flux accumulation is the same, but the longer MR simulations become
magnetically arrested to larger radius, $r\sim18r_g$ in MADfMR
compared to only $r\sim15r_g$ in MADfHR. A difference of only $\sim0.5\%$
in radiative efficiency between MADiMR and MADiHR indicates very good
convergence in our quantity of interest.  We find only small (typically $\sim 10\%$
or less) 
differences between our MR and HR simulations for the other quantities
reported which characterize the accretion flow. For instance, the jet power
differs by only $\sim 50\%$ between MADiMR and MADiHR, unsurprising since 
$\Upsilon\approx5$ for both. Low
resolution `testing' simulations were largely in qualitative
agreement, but with some $Q$'s indicating lack of convergence.

Figs.~\ref{fig:MADiHRvst} and \ref{fig:qsandalphas} show values of
the $\alpha_\mathrm{b}$ viscosity parameter, dominated by the magnetic
stress. We find $\alpha_\mathrm{b}\sim 0.5$, as consistent with
resolved disks in prior MAD simulations.  The effective viscosity
$\alpha_\mathrm{b,eff}$, corresponding to a measurement of the actual
inflow rate indicated by an $\alpha$ parameter, is much larger than
$\alpha_\mathrm{b}$ due to enhanced inflow caused by external torques
driven by low-density highly-magnetized material that introduces
large-scale stresses on the more dense inflow.

\subsection{Power-law Fits for Radial Dependence}
\label{sec:results:vsr}

\begin{figure}
\centering
\includegraphics[width=3.35in,clip]{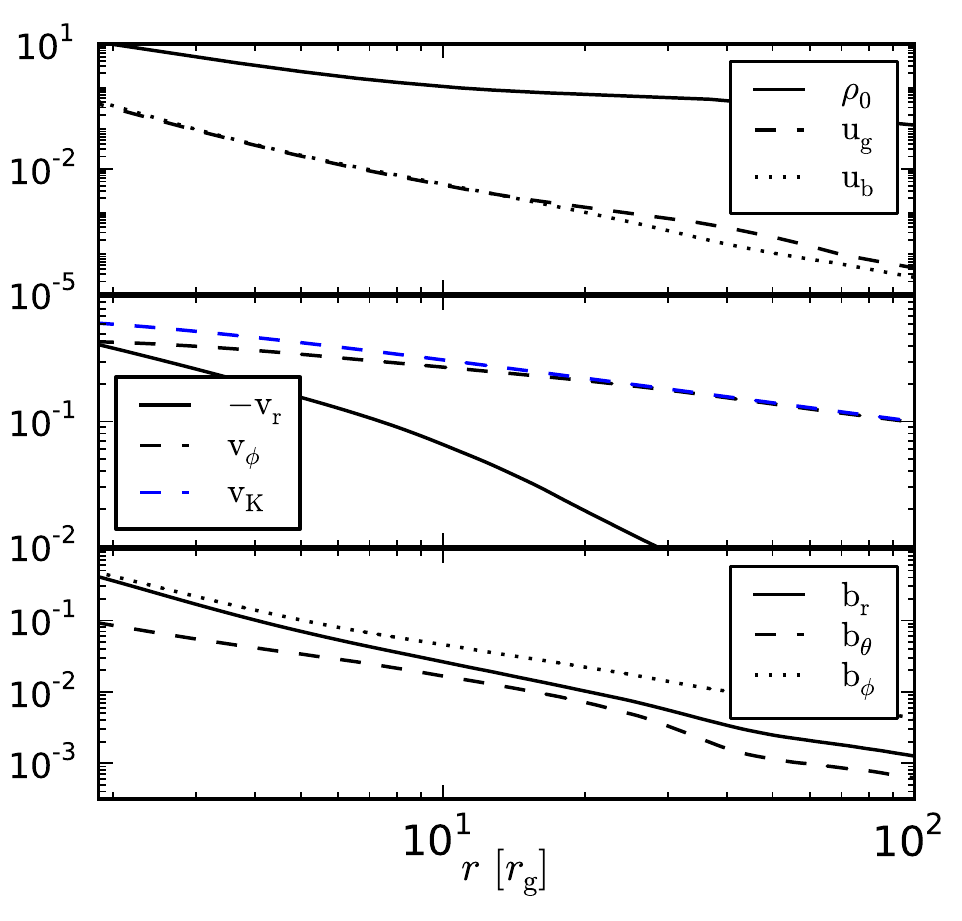}
\caption{Several quantities characterizing the accretion flow in
  MADiHR are plotted as a function of radius. The top panel shows the
  density profile $\rho_0$ (solid), the internal gas energy
  $\mathrm{u_g}$ (dashed), and magnetic energy density $\mathrm{u_b}$
  (dotted). The inner disk, that has reached the MAD state, has
  $\mathrm{u_g}\sim\mathrm{u_b}$ indicating equipartition is achieved
  even within the dense parts of the disk. The middle panel includes
  the disk radial velocity, azimuthal velocity, and Keplerian velocity
  ($\mathrm{v_r}$, solid; $\mathrm{v_\phi}$, black-dashed;
  $\mathrm{v_K}$,blue-dashed, respectively). The final panel shows the
  radial profiles of each component of the comoving magnetic field,
  $\mathrm{b_{r,\theta,\phi}}$, in solid, dashed, and dotted lines
  respectively.  The density is shallower than thicker disk solutions,
  the disk is fairly Keplerian, and the magnetic field falls off as
  expected for accretion disks.}
\label{fig:profiles}
\end{figure}

We now consider radial power-law fits (of the form $f=f_0(r/r_0)^n$)
for quantities associated with the disk structure and flow, plotted in
Fig.~\ref{fig:profiles}, for MADiHR. The fits to disk quantities are
performed as in \cite{2012MNRAS.423.3083M} between radii $r=12r_g$ and
$r=30r_g$ for both our high resolution models.  The profiles for
MADiHR (MADfHR) are $\rho_0\propto r^{-0.4\pm0.1} \:(\propto
r^{-0.8\pm0.2})$, $p_g\propto r^{-1.5\pm0.1} \: (\propto
r^{-1.8\pm0.1})$, $\abs{v_r}\propto r^{-1.6\pm0.06} \:(\propto
r^{-0.8\pm0.2})$, $\abs{v_\phi}\propto r^{-0.38\pm0.01} \:(\propto
r^{-0.49\pm0.01})$ , $\abs{b_r}\propto r^{-1.4\pm0.07} \:(\propto
r^{-1\pm0.1})$, and $\abs{b}\propto r^{-1.2\pm0.05} \:(\propto
r^{-0.89\pm0.02})$.

The MADiHR fits are more representative of a steady-state MAD
solution, because the disk in MADiHR is actually magnetically arrested
over all radii used for fitting and the model is in inflow equilibrium
out to large radii. Our thin disk has a somewhat shallower radial
density profile than prior MAD models, but are otherwise consistent.
To obtain a density profile accurate to large radii, much longer-run
simulations would be required.  The overall fits are consistent with
$p_{\rm g}\propto r^{-1.5}$ and $|b|\propto r^{-1}$ expected for
accretion disks, while the shallow density profile could be connected
with the stronger wind (relative to the jet) compared to thicker MAD
simulations.

%% file: msconclude.tex
\section{Summary and Conclusions}
\label{sec:summary}

We have performed fully 3D GRMHD simulations of radiatively efficient
thin accretion disks seeking to maximize the magnetically-induced
deviations from NT thin disk theory.  We first demonstrated that even
in thin disks, after a time $108,000r_g/c$ and out to $r\sim 17r_g$,
magnetic flux readily advects inward and builds-up to a MAD level (for
which accumulation results in magnetic forces pushing out balancing
against gas forces pushing in).  This occurs despite possible outward
magnetic diffusion through the disk.

Our simulation that is MAD out to $r\sim 35r_g$ after $70,000r_g/c$
has a radiative efficiency of $\eta_{\rm r}\approx 15\%$.  This is
$\approx 80\%$ higher than the standard NT thin disk value of
$\eta_{\rm NT}\approx 8\%$ and is as if the disk were a standard NT
thin disk but with $a/M\approx 0.9$.  BH X-ray binaries have $L/L_{\rm
  Edd}\sim 0.1$ in the high-soft state \cite[]{2011MNRAS.414.1183K},
for which the NT model gives $H/R\lesssim 0.02$ near the BH.  Prior
non-MAD simulations suggest deviations scale with $H/R$
\citep{2010MNRAS.408..752P}, so typical thin MAD deviations might be
$\sim 20\%$.  Such a magnitude of deviations could slightly impact BH
spin measurements, such as in \citet{2011ApJ...743..115N} who found
$6\%$ deviations lead to a BH with $a/M=0$ giving a spectrum like NT
but with $a/M\approx 0.3$.  \citet{2012MNRAS.424.2504Z} found that
emission from near the ISCO is mostly high-energy, but more physics
(e.g., dissipation-radiation energy balance, optical depth
effects, BH photon capture) is required for BH spin fitting
\citep{2011MNRAS.414.1183K}.  We plan to obtain spectra via
post-processing \citep{2015MNRAS.451.1661Z}. Also, clumpy rest-mass 
distribution in thin MAD flows may help to explain spectra from AGN
\citep{2011ApJ...727L..24D} and BH X-ray binaries
\citep{2012MNRAS.426L..71D}.

It is not a new idea that X-ray binaries in the low-hard (LH) state 
supply enough magnetic flux to the BH to enable BZ-type powering
of an observable jet by being geometrically thick, and that the jet 
quenches when the disk collapses to become thin in the high-soft state. 
For instance, plunging material inside the ISCO may enhance the flux 
on the BH in the low-hard state only, even in sub-MAD disks \citep{rgb06}. 
However, these ideas generally propose transition mechanisms in which the jet is 
either powered or not, depending on disk state. Our model does not require
a special mechanism for the LH state. 

Radiatively inefficient flows have $H/R\approx 0.4$
\citep{2012MNRAS.426.3241N}. Our fitting formula Eq.~(\ref{etajfit})
suggests that jet power in the MAD state can drop by $\gtrsim
(0.4/0.02)^2\sim 400$ (or drop by $\gtrsim (0.4/0.01)^2\sim 1600$ for
4U 1957+11 if at $L/L_{\rm Edd}\approx 0.06$) when undergoing
hard-to-soft state transitions. This implies that the observed jet
quenching in BH X-ray binaries \citep{2011ApJ...739L..19R} can occur
despite the presence of a large-scale MAD-level magnetic field in both
disk states.  Such hard-to-soft transitions may also be applicable to
tidal disruption events at late times \citep{2014MNRAS.437.2744T}.
The MAD state could be necessary to explain powerful jet systems
\citep{2014Natur.510..126Z}, while the thermodynamically determined
disk thickness may also play an important role in setting just how
much magnetic flux a MAD has.  Even thinner disks than studied here
would be implied to have $\Upsilon\propto H/R$ and so could have much
weaker magnetic flux and much weaker jets rather than being
independent of the disk state.